\documentstyle[fleqn]{tp}
\input epsf

 at 11truept
\font\tenrm    = cmr10 at 10truept
 at 10truept
						
\font\eightrm  = cmr6 at 8truept

 at 11truept

 at 8truept

 at 12truept                     
 at 11truept

 at 12truept			
 at 11truept

 at 12truept	              
 at 11truept

 at 11truept

 at 11truept

 at 12truept		      
 at 11truept

 at 14.4truept         
 at 12truept         

\def\tenpoint{ \def\rm{ \fam0 \tenrm }      
	\textfont0=\tenrm \scriptfont0=\sevenrm \scriptscriptfont0=\fiverm
	\textfont1=\teni  \scriptfont1=\seveni  \scriptscriptfont1=\fivei
	\textfont2=\tensy \scriptfont2=\sevensy \scriptscriptfont2=\fivesy
	\textfont3=\tenex \scriptfont3=\tenex   \scriptscriptfont3=\tenex
	\textfont\itfam=\tenit      \def\it{ \fam\itfam \tenit}
	\textfont\slfam=\tensl      \def\sl{ \fam\slfam \tensl}
	\textfont\ttfam=\tentt      \def\tt{ \fam\ttfam \tentt}
	\textfont\bffam=\tenbf      \scriptfont\bffam=\sevenbf
	\scriptscriptfont\bffam=\fivebf \def\bf{ \fam\bffam \tenbf}
 	\normalbaselineskip=11.2truept
        \let\sc=\eightrm  \let\big=\tenbig
 	\normalbaselines \rm      }

\def\newpage{\vfill\eject}

\def\apj{ApJ}

\def\aa{A\&A}
\def\mnras{MNRAS}

\newcount\itemno
\def \ino         { \the\itemno\global\advance\itemno by 1 }

\def \mpc	{{\rm\ Mpc}}

\def \kms       {\hbox{ km s$^{-1}$}}

\def \eg	{\hbox{\it e.g.},\ }

\def \etal 	{{\it et al.}\ }

\def \Ho	{{\rm H_{o}}}

\def \kev	{{\rm\ keV}}
\def \msol	{{\rm M}_\odot}
\def \hinv     	{\hbox{$\, h^{-1}$} }

\def \rms	{{\it rms~} }

\def \xray	{\hbox{X--ray} }
\def \se	{\!=\!}
\def \ssim	{\! \sim \!}
\def \sims	{ \sim \!}

\def \spropto	{\! \propto \!}

\def\\{\hfil\break}
\def\spose#1{\hbox to 0pt{#1\hss}}
\def\lta{\mathrel{\spose{\lower 3pt\hbox{$\mathchar"218$}}
     \raise 2.0pt\hbox{$\mathchar"13C$}}}
\def\gta{\mathrel{\spose{\lower 3pt\hbox{$\mathchar"218$}}
     \raise 2.0pt\hbox{$\mathchar"13E$}}}

\def\today{\ifcase\month\or January\or February\or March\or April\or 
    May\or June\or July\or August\or September\or October\or 
    November\or December\fi \space\number\day, \number\year}

\def \logTd6 {\hbox{log$( T/6 \kev)$} }

\def \xray {\hbox{X--ray} }

\def \rtwoh {\hbox{$r_{200}$}\ }

\def \tcdm {$\tau$CDM\ }

\def\etal{{\it et\thinspace al.\ }}

\begin{document}

\twocolumn[
\title{Simulating Large-scale Structure}
\author{August E. Evrard\\
{\it Physics Department, University of Michigan, 48109-1120 USA}}
\vspace*{16pt}   

ABSTRACT.\
After two decades of direct dynamical simulation of large--scale
structure in the universe, 
it is safe to say the subject is now ``mature''.  Still,
there are parts of the problem that are less well developed than
others.  In general, the collisionless dynamics of the dark matter 
component is better understood than the collisional gas dynamics of 
the baryonic component.  In situations where the gas dynamics is
relatively simple, such as the Lyman--$\alpha$ forest and the
intracluster medium in X--ray clusters, our ability to reproduce
observational data has evolved rapidly, and the interpretive and 
predictive power of such experiments should now be taken seriously.  
A comparison of twelve gas dynamic codes to the problem of forming 
a single X--ray cluster shows that numerical inaccuracies are modest 
(typically below ten percent), leaving missing physics as the main
source for large systematic differences between theory and observation.  
Galaxy formation, being more complex, is farther behind in its
development, but simulations capable of resolving the morphological
range of the Hubble sequence in cosmologically interesting volumes may
be just around the corner.

\endabstract]
\markboth{A.E. Evrard}{Simulations of LSS}

\small

\section{Introduction}

Building a universe in the laboratory is a daunting, if perversely
inviting (think of the funding levels!), proposition.  
Fortunately for astrophysicists, there are 
interesting physical questions about the universe that can be
addressed without the creation of a {\sl bona fide\/} mock-up 
--- a reasonable facsimile will do.  How ``reasonable'' the 
facsimile needs to be is
dependent on the level of detail of the questions being posed and the
complexity of the physics driving the processes at hand.  

In this contribution, I will briefly review the status of attempts 
at producing virtual facsimiles of our real universe through direct 
numerical simulation.  In particular, I'll focus on the formation of 
large--scale structure (LSS), where ``large'' means galactic scales
and upward.  Sub--galactic scale
evolution, notably the Lyman--$\alpha$ forest, is reviewed by 
Weinberg \etal in this volume.  The reader will note a general theme 
which is both historical and epistemological; ``simpler'' parts of the 
problem which have been worked 
on for the longest time are better understood than the more difficult
aspects which are only now receiving careful attention.  

\section{N-body Models of LSS}

On it's largest scales, structure in the matter component of the
universe is driven by gravity.  The first attempts at direct N-body
modeling of the gravitational instability process in a volume of
cosmic scale (Aarseth, Turner \& Gott 1979) revealed morphological 
features --- clusters, walls and voids --- characteristic of the 
observed large--scale galaxy distribution.  Future experiments enlarged the
simulated volumes and increased the resolving power, leading to the 
current view of an evolving ``cosmic web'' of dark matter as seen
in the images from the set of $256^3$ particle Virgo simulations
(Jenkins \etal 1998) shown many times at this meeting. 

Since 1970, advances in computing technology --- faster processors,
more memory, better algorithms --- have allowed the typical number of
particles in cosmological simulations to increase by a factor of 100
every decade.  The increased dynamic range has been exploited in two
orthogonal directions which can be roughly characterised as 
increased volume at fixed resolution (minimum 
mass/length scales) or increased resolving power (smaller minimum
mass/length scales) at fixed simulated volume.

 \begin{table*}
 \caption[]{Hubble Volume Models.  Periodic cubes of side $L$ are 
 simulated using CDM fluctuation spectra with shape parameter
 $\Gamma \se 0.21$ and normalizations $\sigma_8$ listed.  The redshifts 
 correspond to look--back epochs along the box diagonal
 $z_{\sqrt{3}L}$, box side length $z_L$ and half-length $z_{L/2}$ for the
 light--cone datasets. }
 \smallskip
  \centering
   \begin{tabular}{l|*{7}{c}}
     \hline
     Model & $\Omega_m$ & $\Omega_\Lambda$ & $\sigma_8$ &
     $L$ & $z_{\sqrt{3}L}$  & $z_L$ & $z_{L/2}$ \\
     \hline
     $\tau$CDM    & 1.0 & 0.0 & 0.60 & 2000 & 4.6 & 1.3 & 0.45  \\
     LCDM         & 0.3 & 0.7 & 0.90 & 3000 & 4.4 & 1.5 & 0.58  \\
     \hline
   \end{tabular}
 \label{tab:1}
 \end{table*}

\subsection{The Hubble Volume Project}

The natural progression toward simulating ever larger volumes has a 
practical endpoint --- the entire visible universe.  Though past
experimenters could have arbitrarily enlarged their volumes to
encompass the Hubble length $c \Ho^{-1} \se 3000 \hinv \mpc$, 
the lack of ability to 
resolve structure out of the linear regime of fluctuations 
left the exercise empty.  Why simulate an analytic system?  

Parallel computers, with their large pools of multi--processor memory
capable of supporting very large $N$ calculations, have changed the 
situation dramatically.   Using 32--bit numerics, the phase space position
plus an index (useful in post-processing) for $10^9$ particles
requires 28 Gb of memory.  A parallel machine with 512 nodes each with
128 Mb of memory offers 65 Gb, enough for $10^9$ particles with space
left over for additional large arrays needed for bookkeeping and
gravitational potential calculations.  

What will a billion\footnote{pardon the Franco-American lingo.} 
particles do for you?  Suppose we ask that the
particle mass be sufficiently small that the Coma 
cluster (mass $\sim 1.1 \times 10^{15} \hinv \msol$) be resolved by  
500 particles.  Then the mass associated $10^9$ particles, $M_{tot} \se
2.2 \times 10^{21} \hinv \msol$, would fill cubes of length $2000
\hinv \mpc$ for $\Omega_m \se 1$ and  $3000 \hinv \mpc$ for $\Omega_m
\se 0.3$.  Since the diagonals of these cubes contain independent information
on a length scale exceeding the Hubble length, they have a legitimate 
claim to be considered simulations of the ``Hubble Volume''.  

Simulations using a billion particles in such Gpc sized volumes have now
been performed by the Virgo Consortium\footnote{see http://star-www.dur.ac.uk/ 
\~{ }frazerp/virgo/virgo.html} using a message--passing version of the Hydra
N--body code (MacFarland \etal 1998) run on a 784 processor 
SGI/Cray T3E at the Rechenzentrum Garching.  The runs each took roughly 5
days of cpu time on 512 processors (equivalent to 7 years of
single--cpu computing) and each generated $\sim 200 Gb$ of output data.  

With such a large simulated volume, the traditional method of recording 
images of the mass distribution at fixed intervals of world time is of
little value for addressing questions of observational significance.
Observational data comes to us from our past light--cone, so
generating output along the past light--cone of artificial
``observers'' in the simulated volume is a natural solution.  The
Hydra code was modified to propagate output filters through the
volume, creating ``surveys'' with different amounts of ``sky
coverage'' to the depths listed in Table~1.  

Details of the geometries of the light cone output 
is available on the web\footnote
{http://www.physics.lsa.umich.edu/ hubble-volume.  Add
/lightcones.htm to this address to get directly to the Hubble tie image.}, and 
a gif image of the ``Hubble tie'', a $40 \hinv \mpc$ thick slice 
along the diagonal wedge survey of the \tcdm model can also be found 
there.  This image, displayed in large format in the foyer of the MPA
building during the meeting, reveals clearly the emergence of
large--scale structure in the universe from $z \simeq 5$ to the
present.  

\begin{figure}
\vskip -2.4truecm
\epsfxsize 10.0truecm
\epsfysize 10.0truecm
\hbox { \hskip -2.5truecm \epsfbox{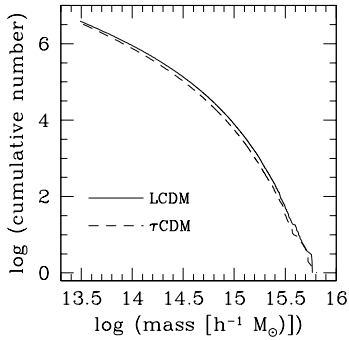} }
\vskip -2.6truecm
\noindent 
\caption{ Cumulative number of clusters identified in the Hubble
Volume models at the final epoch.}  
\label{fig:massftn}
\end{figure}

Figure~\ref{fig:massftn} shows the number of clusters found in each of
the simulated models at the present epoch.  The shear number of
collapsed objects --- roughly 3000 clusters at least as massive as
Coma and nearly one million clusters resolved by more than 32
particles --- allows detailed investigation of statistical questions
such as the behavior of the cluster--cluster correlation function with
cluster richness (Colberg \etal 1998).  

Very rare events can also be studied.  In the \tcdm model at the present
epoch, the largest
group identified with a standard percolation technique lies 
in a region where there are four clusters each more massive 
than Coma lying within a cube of side $20 \hinv\mpc$!  
No obvious counterpart in our local universe comes to mind (the Great
Attractor region may come close), but future deep X--ray,
Sunyaev--Zel'dovich and gravitational lensing studies 
should discover such rarities.

\begin{figure}
\vskip -0.3truecm
\epsfxsize 14.0truecm
\epsfysize 14.0truecm
\hbox { \hskip -5.truecm \epsfbox{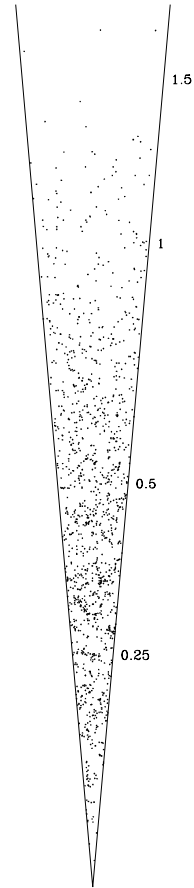} }
\vskip -1.3truecm
\noindent 
\caption{ Positions of clusters in the 100 sq deg diagonal wedge 
light--cone survey from the \tcdm model.  Numbers indicate redshift. }
\label{fig:wedge}
\end{figure}

A view of clusters resolved by at least 32
particles ($M \ge 7.0 \times 10^{13} \hinv \msol$) and 
located in the $10 \times 10$ square degree light--cone
survey along the cube diagonal of the \tcdm run is shown in
Figure~\ref{fig:wedge}.  The plot shows positions in the background 
Robertson--Walker metric (the space of the computation) and the 
linear extent of the
image is 2400 \hinv \mpc.   All the spatial information in the image
is formally independent; no periodic replications are used to generate
the map. 

\begin{figure}
\vskip -1.3truecm
\epsfxsize 11.0truecm
\epsfysize 11.0truecm
\hbox { \hskip -2.5truecm \epsfbox{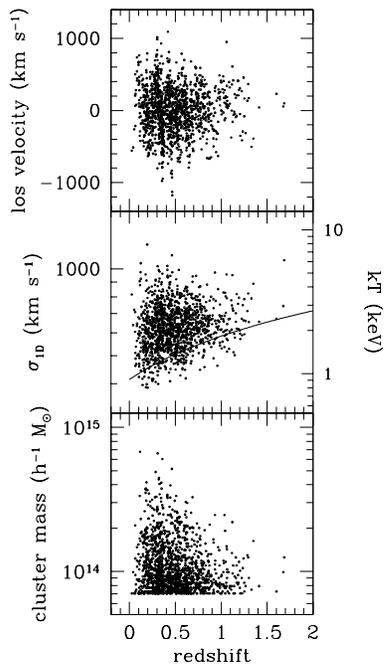} }
\vskip -1.0truecm
\noindent 
\caption{Characteristics of the 1313 clusters displayed in
Figure~\ref{fig:wedge}.   The minimum mass of $7 \times 10^{13} \hinv
\msol$ corresponds to 32 particles.  See the text for discussion. }
\label{fig:TMV_z}  
\end{figure}

A total of 1313 clusters exist in the field, the
furthest located at redshift $z \se 1.68$.  The emergence of clusters
at $z \lta 1$ is evident from the enhancement in number density toward
the vertex.  Hints of the filamentary large--scale matter network are
visible at moderate redshift, but at high redshift the thickness of
the wedge approaches $200 \hinv \mpc$ and projection effects blend
together 
the coherent features which have characteristic length scale $\sim
40 \hinv \mpc$.   

Basic observables of this cluster population are displayed 
in Figure~\ref{fig:TMV_z}.  The panels show the line--of--sight
velocity of the cluster center of mass, the one--dimensional velocity
dispersion $\sigma_{1D}$ and the cluster mass identified using a
friends--of--friends percolation algorithm with linking length 0.2
times the mean interparticle separation in the comoving metric.  

A surprising result from this figure is the fact that the cluster with the 
third highest velocity dispersion in the sample is also the most
distant object at $z \se 1.68$.  The one-dimensional velocity dispersion of
$\sigma_{1D} \se 1072 \kms$ translates into an intracluster medium
temperature $kT \se 6.3 \kev$, assuming a ratio of specific energies 
$\sigma_{1D}^2/(kT/\mu m_p) \se 1.17$, (Frenk \etal 1998).   The
thermal Sunyaev--Zel'dovich (SZ) effect from such a cluster should be
readily observable (see \S3 below).  The existence
of such a potential well at this redshift provides hope that such
targets will begin to appear in deep observational searches for
clusters using SZ, weak gravitational lensing, and even
direct \xray detection (which must fight $(1+z)^4$ surface brightness
dimming).  A few examples of clusters with $z > 1$ have emerged,
including a system discovered optically, CIG J0848+4453, lying at $z
\se 1.273$ with a velocity dispersion $\sigma_{1D} \se 700 \pm 180
\kms$ based on 8 galaxy spectra (Stanford \etal 1997).  Deep ROSAT HRI
images of regions centered on high redshift radio galaxies have also 
provided distant cluster candidates, among them 3C294 at $z \se 1.786$
(Dickinson, private communication).  

The upper panel in Figure~\ref{fig:TMV_z} shows the line--of--sight
(los) mean peculiar velocity  $v_z$ of the clusters as a function of redshift.
This quantity is observable via the kinematic SZ effect, which scales
linearly with $v_z$, but the expected amplitudes are small.  Even for
a cluster with $v_z \se 1000 \kms$, the kinetic amplitude is smaller
than the thermal effect by about a factor 10 (see Birkinshaw 1998).  
Figure~\ref{fig:TMV_z} displays a decrease in the characteristic 
los velocity with redshift, as expected from linear theory, indicating
the kinematic SZ effect as a poor choice for detecting high-$z$
clusters.  

Of course, the fact that these processes locally distort the cosmic 
microwave background means that the CMB is the ultimate search engine for 
high redshift clusters.  The future MAP and Planck missions, which will
provide nearly all-sky coverage, will see bright clusters as foreground
hot and cold spots, depending on wavelength, subtending roughly arcmin 
angular sizes.  The lightcone data from the Hubble Volume will be used
as a template for such effects, with gas physics being introduced ``by
hand'' into the potential wells in a model dependent manner (Kudlicki
\etal, in preparation).  

The existence of a hot cluster at high redshift is assisted by the
fact that the virial temperature at a fixed mass scales as $T \spropto
M/r \spropto \bar\rho^{1/3}(z) \spropto 1+z$ (Kaiser 1986).  This relation 
is shown in the middle panel of Figure~\ref{fig:TMV_z} for 
clusters at the 32 particle
limit assuming $\sigma^2_{1D} \se 1000 (M/10^{15}\hinv
\msol)^{2/3} (1+z)$ (\eg Bryan \& Norman 1998).  Clusters at the low 
mass cutoff follow this mean relation well (with anticipated 
$\ssim 10\%$ scatter). 

The $z \se 1.68$ cluster actually has mass a factor two above 
the 32 particle limit.  Is a $1.4 \times 10^{14} \hinv \msol$ 
cluster expected in a survey of this size at this redshift?  
From Figure~\ref{fig:TMV_z}, it is evident that
this object is a rare occurrence, but it does not seem pathologically
so.  The mass spectrum at all redshifts has a naturally ragged 
high mass envelope, and the $z \se 1.68$ object appears a reasonable 
extension of the trend exhibited at lower $z$.  

More quantitatively, 
one can use the traditional link
between linear fluctuation amplitude and collapse epoch $\delta(z) \se
1.68 (1+z)$ along with the employed \tcdm normalization of the initial
Gaussian fluctuations to estimate that this object corresponds to roughly a
$5\sigma$ upward density fluctuation.  Now $5\sigma$ is certainly
rare --- the tails contain $p \se 2 \, {\rm erfc}(5) = 6 \times 10^{-7}$ of the
Gaussian pdf --- but countering this rarity is the large number of
independent samples of 64 particle masses available in the Hubble
Volume simulation $N \se 10^9/64 \se 1.5 \times 10^7$.  The product $Np$
implies there should be about 10 such clusters in the entire
simulation volume.  Since
the survey volume between $z \se 1.5$ and $z \se 2$ in
Figure~\ref{fig:wedge} represents about $1\%$ of
the whole, it's a bit lucky (1 in $\sims 10$) but not crazy
that one such cluster appeared in the survey volume.  

\subsection{Halo Structure}

One can use the increased dynamic range from parallel computers to
probe the internal structure of individual objects such as 
clusters to higher density contrasts (Moore \etal 1998; Dubinski
1998).  Compared to using the increased
particle number to ``buy volume'', this is a technically more
demanding task.   Since the gravitational dynamical time scales as 
$(G \rho)^{-1/2}$, resolving structure to higher densities requires
integrating particle orbits using shorter timesteps and over more
dynamical times.  

Moore \etal (1998) break new ground with simulations in which 
a single cluster is resolved with over 3 million particles interior
to the virial radius (conventionally defined as $r_{200}$, the radius of the
sphere within which the mean interior mass density is 200 times 
the critical density).  With force resolution of 5 and 10 kpc in two
separate runs, they approach three orders of magnitude dynamic range
in linear scale within the virial radius of 2 Mpc for their simulated
``Virgo'' replica (the name refers to the observed cluster, not the
consortium).  They find the inner portion of the density profile to 
approach a power law $\rho(r) \spropto r^{-1.4}$, steeper than 
the asymptotic slope of $-1$ predicted by the model of Navarro, Frenk
\& White (1997 and references therein; hereafter NFW)  
\begin{equation}
\frac{\rho(r)}{\rho_c} = \frac{\delta_c}{cx (1+cx)^2} .
\label{eq:nfw}
\end{equation}
Here $x\se r/r_{200}$ is the scaled radius, $c$ is the concentration
parameter --- the sole fitting parameter --- and $\delta_c \se (200/3) 
c^3/[\ln(1+c)-c/(1+c)]$ ensures consistent normalization of the
profile at $x=1$.  

Moore \etal find that the NFW form does not provide an acceptable fit
across the entire range of densities probed by their experiment;  for
any value of the concentration parameter, there remains structure in
the residuals about the fit.  The interior portion of the profile
appears to be driving the conflict.  Fits to lower resolution runs
are acceptable, but the inner part of the profile appears
unwilling to cooperate and roll over to the $r{-1}$ form expected from
equation(\ref{eq:nfw}).  

Population aspects of The NFW profile were presented at this meeting
by Jing (these
proceedings) who examined a large body of halos from a set of 16.8
million particle runs performed on a parallel supercomputer with 
16 fast vector processors.  Jing examines complete samples of halos
within the simulations and generates impressive statistics.  A single run
contains 300 to 500 halos resolved by more than 10,000 particles.  
He finds that a significant fraction of objects have density profiles 
that are poorly fit by equation(\ref{eq:nfw}). 
 
At first clance, this result is not surprising, since the studies 
by NFW selected {\sl against\/} halos which were dynamically active
while Jing's study includes all objects.  However, what is surprising 
is the size of the sample which is not well fit by the NFW form;  
over a third of the clusters in the sample exhibit maximum fractional 
deviations of at least $35\%$ from the NFW form (using 10 bins  
per decade in radius).  The best fits for this fraction yield a
distribution of concentration parameters quite different from the 
remaining clusters in the sample --- the peak concentration is lower
by a factor two and the width is broader by a similar factor.  

What does this all mean for the NFW profile?  An obvious lesson to
learn from these new studies is that one must
exercise care in application of the density profile in 
equation(\ref{eq:nfw}).   It is not ``universal'' in the sense of
describing any single halo drawn from a random cosmology (nor was it
ever described as such by NFW, who point out its applicability to
relaxed systems).  

For all its newfound faults, all experiments done to date confirm that
the form of equation(\ref{eq:nfw}) does a remarkable job of describing a 
significant fraction of the mass density profiles in the majority of 
collapsed halos.  
The NFW profile remains the {\sl de facto\/} standard 
analytic form for halo density profiles, and it will remain so until
future studies provide well calibrated modifications.  
With apologies to Ivan King, a return to truncated isothermal 
spheres (King 1962) would be a step in the wrong direction.  

One of the practical ramifications of Jing's work is that 
observers cannot rely on 
observation of the mass profile in a single cluster (\eg Tyson,
Kochanski \& Dell'Antonio 1998) 
to place strong constraints on cosmology.  Samples of 
ten or so clusters would provide interesting constraints, 
particularly if the defining sample criteria were tailored to select
against objects undergoing significant mergers.  X--ray imaging and
spectroscopy could provide the clues necessary for such sample
selection, but this requires understanding of the dynamical 
behavior of the intracluster medium in clusters.

\section{``Dissipationless'' Gas Dynamics : X--ray Clusters}

The hot intracluster plasma that permeates clusters of galaxies
represents the next most important matter constituent of clusters after dark
matter.  The fact that this intracluster medium (ICM) outweighs the
matter associated with cluster galaxies is one of the legacies of the
ROSAT satellite mission.  Based on the ROSAT all-sky survey imaging of
Briel, Henry \& B\"ohringer (1992) and galaxy photometric data of Godwin,
Metcalfe \& Peach (1983), White \etal (1993) find a gas--to--galaxy 
mass ratio $M_{gas}/M_{gal} \se (5.5 \pm 1.5) h^{-3/2}$ for the Coma
cluster within an Abell radius $r_A \se 1.5 \hinv \mpc$.   This
translates to a factor 10 for $h \se 0.65$.  

The dominance of the ICM in the baryonic budget of clusters has a
number of implications.  From a modeling perspective, it supplies
ammunition to justify the assumption
that interaction between galaxies and the ICM is a
``higher--order'' effect, at least in very rich clusters such as Coma.  
This leads to the basic treatment of the dynamical and thermodynamic 
behavior of the ICM as being driven by 
the gravitational evolution of the dominant, dark matter.  A complex
network of shocks of varying strengths develop naturally in the 
merger/accretion process and the combined action of these shocks heats
the gas by thermalizing the infall energy of the gravitational
clustering process.

\begin{figure*}
\vskip -0.5truecm
\epsfxsize 10.0truecm
\epsfysize 10.0truecm
\hbox { \hskip -0.1truecm \epsfbox{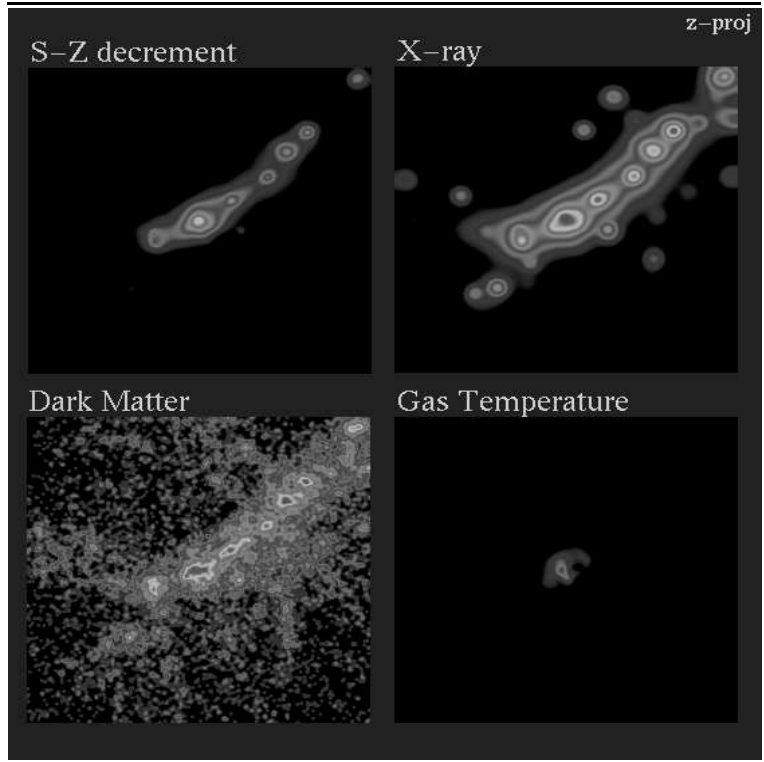} }
\vskip -0.0truecm
\noindent 
\caption{Views of a cluster at $z \se 1.24$.}
\label{fig:4pack_z1.2}  
\end{figure*}

\begin{figure*}
\vskip -0.5truecm
\epsfxsize 10.0truecm
\epsfysize 10.0truecm
\hbox { \hskip -0.1truecm \epsfbox{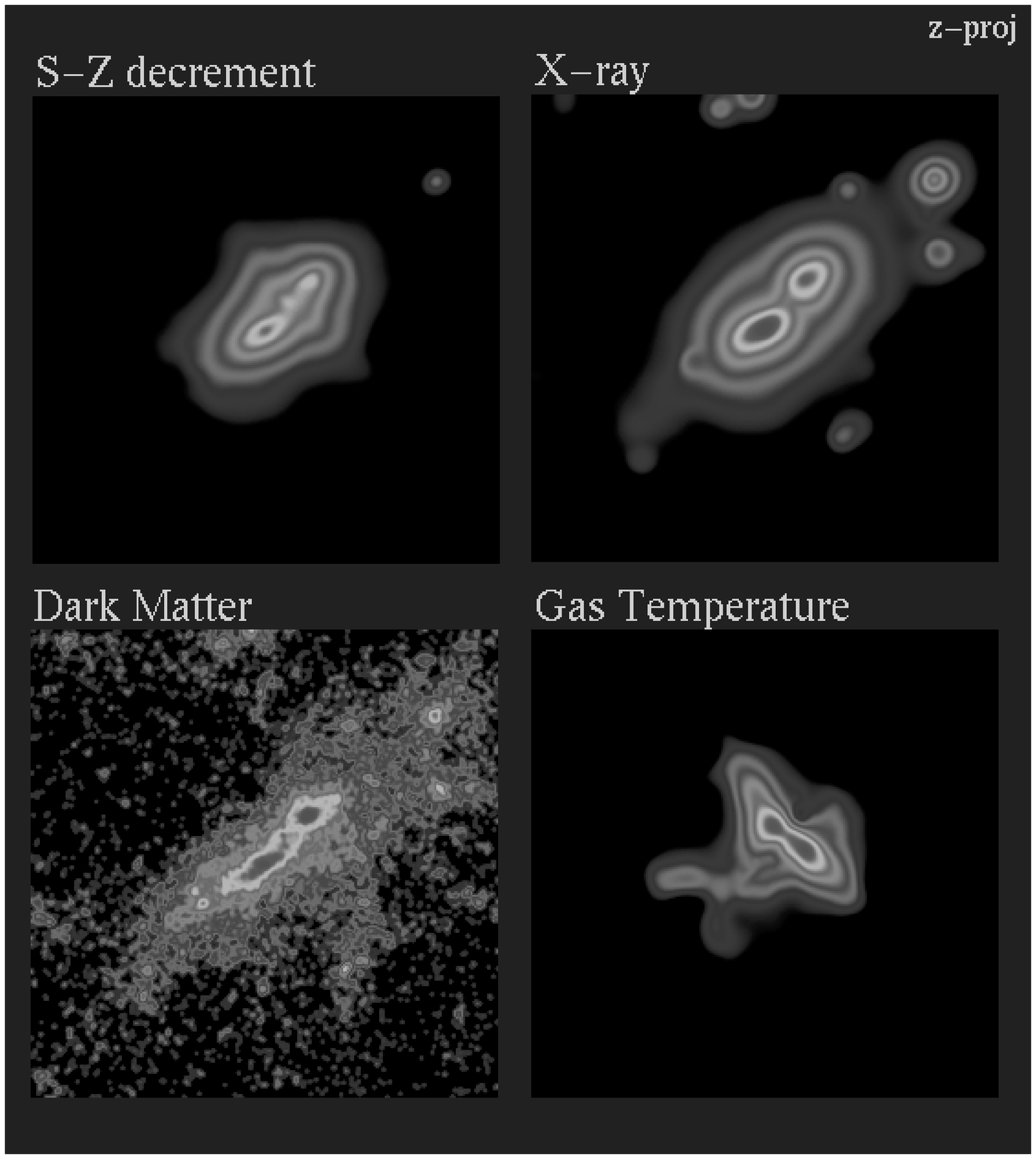} }
\vskip -0.0truecm
\noindent 
\caption{Views of a cluster at $z \se 0.6$.}
\label{fig:4pack_z0.6}  


\vskip 0.5truecm
\epsfxsize 10.0truecm
\epsfysize 10.0truecm
\hbox { \hskip -0.1truecm \epsfbox{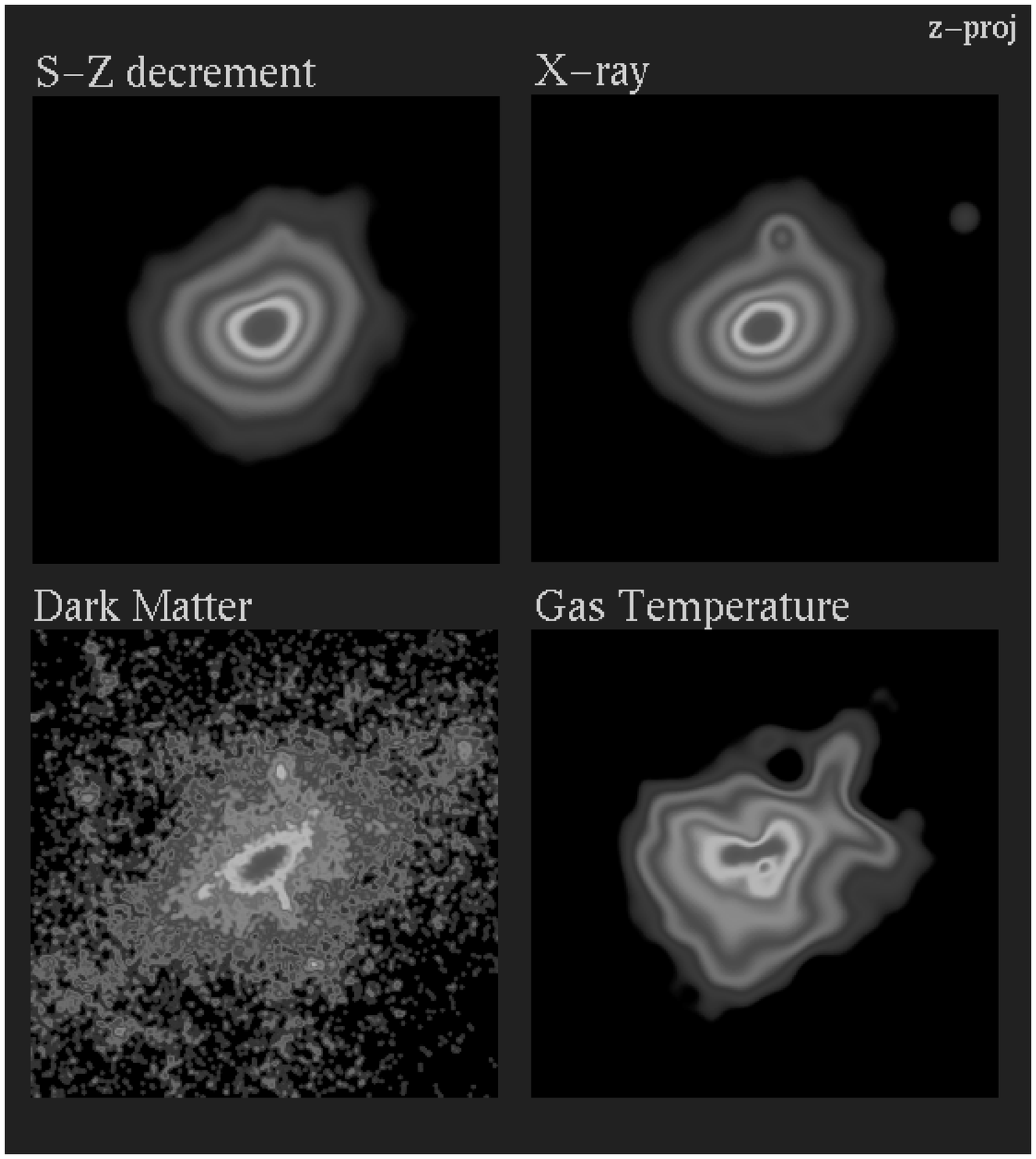} }
\vskip -0.0truecm
\noindent 
\caption{Views of a cluster at $z \se 0$.}
\label{fig:4pack_z0}  
\end{figure*}

In the nearly decade since this process was first examined using 
self--consistent collisional gas and collisionless dark simulations 
matter (Evrard 1990a,b), much has changed but much has stayed the
same.  One of the things that has stayed the same is efficiency of 
thermalization of the gas, traditionally expressed as 
$\beta \se \sigma^2_{1D}/(kT/\mu mp)$, the ratio of specific energies 
in dark matter and gas.  In a comparison of twelve gas dynamics codes
to the formation of a single \xray cluster (drawn from a standard cold
dark matter model with $10\%$ baryons), Frenk \etal (1998) find 
$\langle \beta \rangle \se 1.17 \pm 0.06$ within $r_{200}$ to be the
mean and standard deviation among the twelve codes, consistent with 
the value $\beta \se 1.2$ found by Evrard (1990a).  A noteworthy
point is that, despite their very different numerical treatments of
shocks, codes of different type --- Eulerian, Lagrangian hybrid ---
produced very similar values of $\beta$. 

A look at the P3MSPH code (Evrard 1988) solution to the comparison
cluster is provided in Figures~\ref{fig:4pack_z1.2},
~\ref{fig:4pack_z0.6} and ~\ref{fig:4pack_z0} which show four
principle cluster observables at redshifts $z \se 1.2, 0.6$ and 0,
respectively.  The four panels show greyscale representations of the
logarithms of the anticipated thermal Sunyaev--Zel'dovich signal (over
2 decades), intrinsic \xray surface brightness (4 decades), gas
temperature (factor 4, from $0.25-1 \times 10^8$ K) 
and dark matter (DM) surface density (2.5 decades).
The gas maps are smoothed on the local SPH kernel scale while the DM
is left unsmoothed to show the graininess in the particle solution.  
A physical region of size $3.9 \hinv \mpc$ is shown at all redshifts.  
An mpeg animation of this four--panel figure is available at 
http://astro.physics.lsa.umich.edu/
evrard/VCL/ccp\_movie.htm.

There are characteristic features in these images
common to all twelve codes in the comparison study.  
At $z \se 1.2$. most of the mass in the region has already collected
into a long, knotty filament which will set the orientation for a
chain of future merger events between the knots.  One can imagine a
network of such filaments covering the sky at these redshift --- 
the ``proto--cluster formation epoch''.  A good portion of the 
absolute scale the SZ decrement shown in the figures, $3.4$ to $340 \
\mu$K, is visible with
current ground based techniques (\eg Carlstrom, Joy \& Grego 1996) and 
the cosmic web that is the large--scale distribution of clusters (Bond
\& Myers 1998) may be revealed by future CMB experiments.  

An interesting set of independently observable features occurs in a 
pre--merger phase displayed in Figure~\ref{fig:4pack_z0.6}.  The
merger event involves a triplet, consisting of a pair of satellites 
(projected out along the
line-of-sight in this image) each with mass in excess of $20\%$ of 
the dominant precursor's mass.  At the time shown, the thermalized gas
envelopes of the triplet are being compressed and mildly shocked. 
The \xray image displays two prominant peaks (the projected pair of
satellites is to the upper right) and a 
strong emission weighted temperature gradient exists along the 
collision axis near its vertex.  This pinching effect comes about 
because the gas at the cores of the precursors has not yet been moved
off their incoming adiabats, while gas trapped near the systems center
of mass is squeezed hard and subsequently strongly shocked.  
The hot pinched spot with adjacent sidelobes is a distinct 
signature of a merger occurring along the line of sight.  

The electron pressure visible in the SZ image has features common to
both the gas density (X--ray) and temperature.  Two peaks are separated
by a pressure ridge  of strongly shocked material perpendicular to 
the collision axis.  Because of the confining ram pressure along the
collision axis, the flow of gas at this time is highly anisotropic.  
Inward flow along the axis feeds outward flow off-axis, and a small
fraction of gas is shed into the voids
which surround the large--scale filament.  The dark matter density
distribution is highly flattened, with two peaks evident as in the
\xray image.   

Observations with the same dynamic range and resolution of these maps
do not yet exist.  In this sense, theory is ahead of the
observation, and one can interpret these maps as 
predictions for upcoming, high resolution ground--based 
and satellite observations.  
The spectroscopic imaging capabilities offered by AXAF will soon begin
probing the line--of--sight velocity structure, and observable velocity
splittings are anticipated (Norman \& Bryan 1998).  

\begin{figure}
\vskip -2.0truecm
\epsfxsize 9.0truecm
\epsfysize 9.0truecm
\hbox { \hskip -2.0truecm \epsfbox{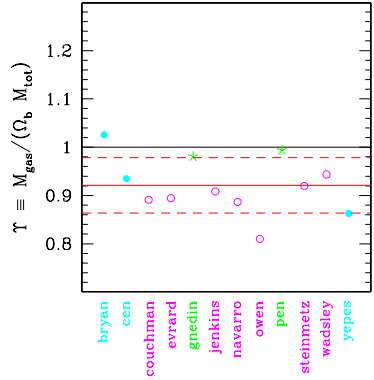} }
\vskip -2.0truecm
\noindent 
\caption{Normalized gas fractions within \rtwoh for the twelve codes
in the cluster comparison study.  Open symbols are Lagrangian (SPH)
codes, filled circles Eulerian and stars hybrid. }
\label{fig:Upsilon}  
\end{figure}

Despite this seemingly active formation history, relatively 
little of the gas that was initially associated with the dark matter 
in the virial region is lost from the system at the present epoch.  
Figure~\ref{fig:Upsilon} shows the normalized gas fraction  
$\Upsilon(r) \se M_{gas}(r)/\Omega_b M_{tot}(r)$ evaluated within
\rtwoh for the twelve clusters simulated in the comparison study.
The solid and dashed lines show the mean and $1 \sigma$ range 
$\langle \Upsilon \rangle \se 0.92 \pm 0.06$ determined from the
experiments.  The Lagrangian codes (open symbols) seem to lose 
somewhat more of the gas than the other code types, but the source of
this difference is not yet understood.

\begin{figure}
\vskip -2.0truecm
\epsfxsize 9.0truecm
\epsfysize 9.0truecm
\hbox { \hskip -2.0truecm \epsfbox{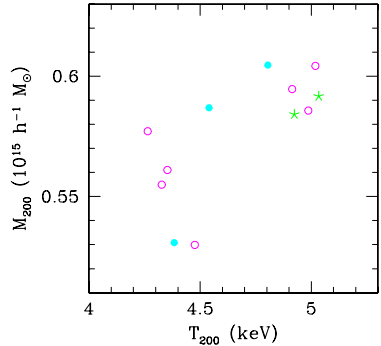} }
\vskip -2.2truecm
\noindent 
\caption{Temperatures and masses from the cluster comparison
study. Note the $20-30\%$ range of the axes. }
\label{fig:T_M}  
\end{figure}

Determination of the normalization of this effect is important for
constraints on the total mass density $\Omega_m$ from the mean cluster 
baryon fraction (White \etal 1993; Evrard 1997).  Another important
factor in that argument is the normalization of the virial 
mass--temperature relation.  Experiments show small scatter about the
virial relation (Evrard, Metzler \& Navarro 1996; Bryan \& Norman
1998; Eke, Navarro \& Frenk 1998), but the zero point remains 
uncertain to perhaps $15 \%$.  A look at the agreement between virial mass
and (mass--weighted) temperature for the cluster comparison study 
is given in Figure~\ref{fig:T_M}.  Hints of the overall
relation $M \spropto T^{2/3}$ in this diagram suggest that part of the
disagreement among codes is due to a lack of strict synchronization in
the cluster's dynamical state.  Small differences in the treatment of 
tidal fields in the linear regime, for example, can be amplified into
large orbital phase differences for infalling small satellites such as
that visible to the north in Figure~\ref{fig:4pack_z0}.  Images based
on the twelve codes' solutions exhibit noticeable differences in the
positions of satellite objects. 

Table~\ref{tab:ccp} summarizes the extent of the agreement among codes
for internal structural properties of the cluster at the final time.
The \rms deviation of spherically averaged, binned data interior to 
the virial radius of the cluster (using 0.2 dex binwidth) is shown.
Agreement in all the listed quantities is better than $10\%$,
except for the bolometric \xray luminosity, which has an \rms of
$23\%$.  The luminosity $\L_X \se \int dV \rho_{gas}^2(r) T^{1/2}(r)$
is particularly sensitive to the core gas structure, and the larger
deviation in $L_X$ reflects the fact that the codes agree less well on
the core structure than they do on the exterior envelope.  

 \begin{table}
 \caption[]{Level of agreement between twelve different cosmological gas
 dynamics codes in the final, virial structure of a single 
 \xray cluster (Frenk \etal 1998).}
 \smallskip
  \centering
   \begin{tabular}{l|*{2}{c}}
     \hline
     Quantity & \rms deviation \\
     \hline
     $\rho_{DM}$ & 0.031  \\
     $\rho_{gas}$ & 0.086  \\
     $\sigma_{DM}$ & 0.046  \\
     $T_{gas}$ & 0.053  \\
     $P_{gas}$ & 0.081  \\
     $L_{X}$ & 0.23  \\
     \hline
   \end{tabular}
 \label{tab:ccp}
 \end{table}

Additional physics such as radiative cooling and magnetic fields will
also affect the core structure, perhaps quite strongly.  These effects
are unlikely to affect the structure of the bulk of the gas in very
rich clusters, but may contribute perhaps tens of percent effects.  Many
effects are only now being considered with careful experiments.  
Dolag, Bartelmann \& Lesch (1998) displayed a poster at the meeting
showing that the effects of magnetic fields on cluster evolution.   
Teyssier, Chi\'eze \& Alimi (1997) include non--equilibrium
thermodynamics and examine differences in electron and ion
temperatures.  Metzler \& Evrard (1997) examine the effects of
galactic winds.  None of this additional physics appears to change 
the overall picture of the ICM presented above, though some 
details will certainly be affected.  
Dissenters point out that a strongly multi--phase intracluster medium
is not formally ruled out (Gunn \& Thomas 1996), but upcoming SZ
maps combined with AXAF and XMM spectroscopic imaging will provide
definitive constraints (Nagai, Sulkanen \& Evrard 1998).  

\section{Galaxy Formation}

\begin{figure*}
\vskip -0.0truecm
\epsfxsize 10.0truecm
\epsfysize 10.0truecm
\hbox { \hskip -0.4truecm \epsfbox{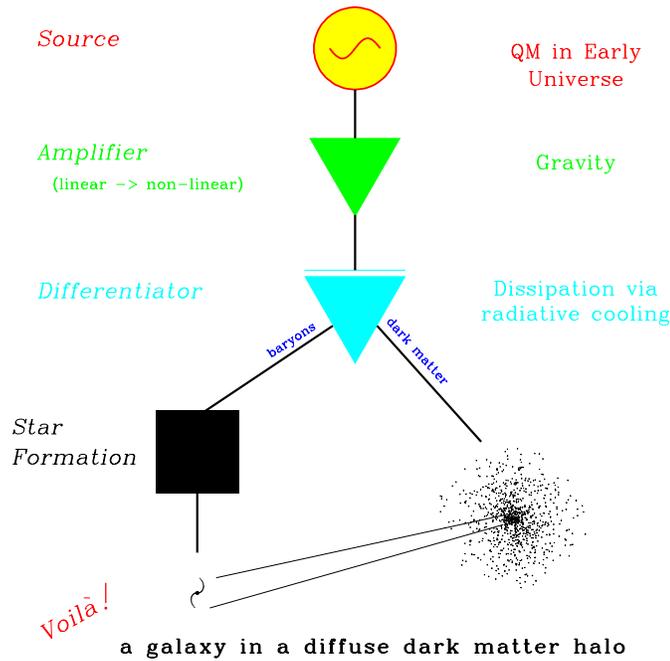} }
\vskip -0.3truecm
\noindent 
\caption{ Flow diagram of the galaxy formation process.}
\label{fig:EE}  
\end{figure*}

Galaxy formation is the perennial frontier in the simulation
business.  As outlined in Figure~\ref{fig:EE}, the 
radiative cooling instability within collapsed, dark matter
potential wells is the essential ingredient which allows the 
baryons to sink to the center, become self--gravitating and, thereby, 
light up (White \& Rees 1978).  Resolving this instability is
numerically  challenging (see the heroic efforts of Abel, Bryan \&
Norman in this volume).   One has the added complication that, even
with a ``perfect'' numerical code, the physics governing star forming
regions is not yet understood.  The problem is formally
ill--posed and solutions will be guided as much by instinct and
intuition as by first principles.  

Radiative cooling of the gas has been enabled in a number of calculations
(Katz \& Gunn 1991; Katz, Hernquist \& Weinberg 1992; 
Evrard, Summers \& Davis 1994; Navarro \& White 1994; Steinmetz \&
M\"uller 1994;  Frenk \etal 1996;  Tissera, Lambas \& Abadi 1997; 
Navarro \& Steinmetz 1997) and a population of cold, condensed baryonic
cores develops within their enveloping halos, largely along the lines 
scripted by the sages twenty years ago.  A number of potentially 
interesting results have emerged, but their robustness remains 
in question.  

The issue of ``biasing'' --- how the phase space structure of galaxies
differs from that of the dark matter --- is of crucial importance to
interpreting data from upcoming deep optical surveys such as the 
Two Degree Field (2dF, see Maddox in this volume) and the Sloan 
Digital Sky Survey (SDSS).  Biasing is a complicated issue because
both gravitational (dynamical friction, merging, galaxy
harassment) and non--gravitational effects (dependence of gas cooling
and star formation rates on large--scale environment) may play equally
important roles.  Direct simulation of cluster environments (\eg Frenk
\etal 1996) in conjunction with semi--analytic approaches which
postulate recipes for star formation and feedback in halos (\eg
Kauffmann, Nusser \& Steinmetez 1997) are beginning to yield clues.

\begin{figure}
\vskip -0.0truecm
\epsfxsize 5.0truecm
\epsfysize 5.0truecm
\hbox { \hskip -0.1truecm \epsfbox{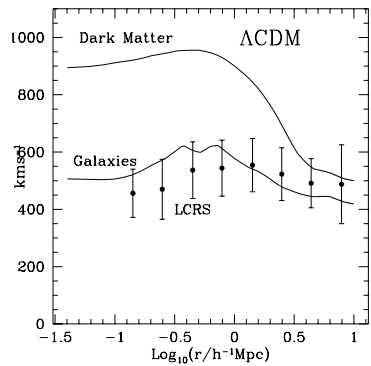} }
\vskip -0.2truecm
\noindent 
\caption{Distribution of peculiar velocities for galaxies derived from
a 16 million particle SPH calculation (Pearce \etal, in preparation). }
\label{fig:sigmagal}  
\end{figure}

An example of a result which may turn out to be robust is shown in
Figure~\ref{fig:sigmagal}.  The figure shows the pairwise peculiar
velocity of galaxies and dark matter as a function of pair separation 
derived from a 16 million particle simulation of an LCDM universe
(Pearce \etal, in preparation).   Galaxies --- identified with cold, high 
density baryonic knots in the calculation --- show a lowered overall
velocity distribution compared to the dark matter and this produces a
much better agreement with observational values derived from the 
LCRS redshift survey (Lin \etal 1996).  Whether or not this agreement
is fortuitous hinges on the sensitivity of this result 
to details such as numerical parameter choices, galaxy sample 
definition (both real and simulated) and the weighting of the velocity
statistic (Davis, Miller \& White 1997).  Still, the agreement between
theory and observation in Figure~\ref{fig:sigmagal} is encouraging and
it may indicate that the current simulations are accurately 
capturing galaxy dynamics within the large--scale web of dominant 
dark matter.  

\section{Summary}

The present state of numerical 
modeling of large--scale structure in the universe is 

{\obeylines \hangindent=1.truecm \hangafter=0
$\bullet$ well understood and ahead of observations of dark matter structure, 
$\bullet$ becoming understood and on par with observations of \xray clusters and the Ly-$\alpha$ forest, 
$\bullet$ poorly understood and far behind observations of the galaxy distribution. 
}

What does the future hold?  We can expect to see parallel
computers used both to expand dynamic range (by allowing bigger calculations) 
and also to explore model parameter space (by allowing large numbers
of smaller calculations).  The number and range of physical processes being 
included in the calculations will certainly increase.  The downside of this
added complexity is that it will naturally introduce additional
sources for numerical error.  Besides highly simplified systems, the
main check on the numerical solutions may remain internal --- 
a ``galaxy comparison project'' analogous to that performed for
clusters is likely to occur in the near future.  
The upside of the added complexity is the improved contact it will 
bring with observational data.  Spectroscopic imaging, broad band
colors, kinematic studies, etc. --- the sky is the target (but 
not necessarily the limit) in the virtual world.

It is easy to foresee the problem of galaxy formation being
scrutinized more thoroughly both from the 
``bottom--up'' (via modeling of absorption line systems and 
high--$z$ progenitors of galaxies) and from the ``top--down''
(via modeling of galaxies in different large--scale environments,
clusters versus the field).  By attacking the problem of galaxy
formation from below and above in this way, we will squeeze out 
the answer to the lingering question ``how do galaxies 
trace the mass'' in the universe.

\section*{Acknowledgments}

This research was supported by NASA's Astrophysics Theory Program 
and by grant AST-9803199 from NSF.  The Max--Planck Society and UK PPARC 
provide primary support for the Hubble Volume Project.  
I am very grateful to my Virgo
collaborators for allowing me to display our joint work in these
proceedings.  Thanks to Tony Banday and the rest of the 
organizers for putting together such an excellent meeting.


\end{document}